\documentstyle[fleqn,epsfig]{article}
\begin{document}
\title{About EAS inverse approach}

\author{S. V.~Ter-Antonyan\footnote{e-mail:samvel@jerewan1.yerphi.am}\\
\emph{Yerevan
Physics Institute, Alikhanyan Brothers 2, Yerevan 375036,
Armenia}}
\date{} 
\maketitle
\begin{abstract}
It is shown that using the observed extensive air shower (EAS) 
electron and truncated muon size spectra at sea level
one can solve the EAS inverse problem - reconstruction of primary
energy spectra and elemental composition, for not more than 2
kinds of primary nuclei.
\end{abstract}
\newpage

\section{Introduction}

High accuracy of modern EAS experiments in the primary
energy region of
$10^{15}$ eV 
[1-4] 
increased a number of 
publications on the
solutions of the
EAS inverse problem, which is reconstruction of
primary nuclei energy
spectra based on the detected EAS parameters at
observation level 
[5-9]. 
However, discrepancies in results continue to grow.
In most of the cases, it is a result of the hidden
experimental systematic errors
due to uncertainties in the response functions of
detectors.
The uncertainty of A-A interaction model at these
energies also contributes
to the discrepancies in primary nuclei energy spectra
results.
However, there are certain publications where the EAS
inverse problem is solved
based on erroneous presuppositions, which of course
only increase the existing discrepancies of data.
\section{EAS inverse problem}
In general, the relation between energy spectra ($\partial
\Im_{A}/\partial E$) of 
primary nuclei ($A\equiv1,4,\dots56$) and measured EAS 
electron and muon size spectra at observation level ($\Delta I/
\Delta N_{e,\mu}^*$) at a given zenith angle $\theta$ is
determined by an integral equation
\begin{equation}
\frac{\Delta I(\theta)} {\Delta N_{e,\mu}^*}=
\sum_{A}
\int
\frac{\partial \Im_{A}} {\partial E}
\frac{\partial K(E,A,\theta)}{\partial N_{e,\mu}^*} 
dE 
\end{equation}
The kernel function of equation (1) is determined as  
\begin{equation}
\frac{\partial K}{\partial N_{e,\mu}^*}\equiv
\int
\frac{\partial W(E,A,\theta)}{\partial N_{e,\mu}}
\frac{dG}{dN_{e,\mu}^*}dN_{e,\mu}  
\end{equation}
where the EAS size spectra $\partial W/\partial N_{e,\mu}$ 
depend on observation level, $A,E,\theta$ 
parameters of primary nucleus and $A-A_{Air}$ interaction model, 
$dG(N,N^*)/dN^*$ is the error function of measurements.\\  
Equation (1) is a typical ill-posed problem and 
has an infinite set of solutions for unknown primary energy 
spectra. However, the integral equation (1) turns to a
Fredholm equation if a kind  of a primary nucleus is defined 
directly in the experiment 
(as it successfully does in the balloon and
satellite measurements \cite{JACEE,RUNJOB} where energy 
spectra for different nuclei are obtained up to $10^{15}$ eV)
or the detected EAS parameters (or combination of 
parameters) slightly depend on primary nuclei \cite{Roma}.
In the last case one evaluate only the all-particle energy 
spectra $\sum_{A} d\Im_{A}/dE$ from equation (1).\\
As a result, the meaning of data presented in the 
\cite{KAS1} and further 
publications 
[12-14]
where authors claim to have 
got the solution of equation (1) for 4 types of primary nuclei, 
is not clear.\\
Let's prove that using the measured EAS electron and truncated
muon size spectra at 3 zenith angular intervals, as its done in
\cite{KAS1}, the Eq. (1) can
have a single solution only for primary flux which consists of
not more than 2 kinds of nuclei.\\

Let the EAS size spectra at observation level KASCADE ($t=1020$
 g/cm$^2$ \cite{KAS1}) be 
described by log-Gaussian distributions with mean
$<N_{e,\mu}(E,A,\theta)>$ and variance 
$\sigma_{e,\mu}^2(E,A,\theta)$. It is known that this 
assumption is well performed at atmosphere depth $t>700$ g/cm$^2$,
primary energies $E>10^5$ GeV and zenith angles $\theta<35^0$.\\ 
Let also the measurements and further evaluations 
of EAS electron and truncated muon sizes
be carried out without errors ($dG/dN^*\equiv\delta(N-N^*)$)
and integral Eq.~(1) include only statistical uncertainties.\\
Then a set of Eq.~(1) transforms into the following:
\begin{equation}
\frac{\Delta I(\theta)} {\Delta N_{e,\mu}}=
\sum_{A}
\int
\frac{\partial \Im_{A}} {\partial E}
\frac{\partial W(E,A,\theta)}{\partial N_{e,\mu}} 
dE
\end{equation}
Let's determine the parameters of distribution functions $\partial
W/\partial N_{e,\mu}$ by the following known empirical expressions: 
\begin{equation}
<N> \simeq a
\Big(\frac{E}{1GeV}\Big)^{b}A^{c}\cos^{d}\theta
\end{equation}
\begin{equation} 
\sigma\simeq\alpha A^{\delta}
\Big(\frac{\ln
(E/1GeV)}{\ln10^6}\Big)^{\varepsilon}\cos^{\rho}\theta 
\end{equation} 
where the values of corresponding
approximation parameters ($a,\dots d,\alpha,\dots\rho$) are
presented in Tables 1,2 and
obtained by
CORSIKA6016(NKG) EAS simulation code \cite{CORSIKA} at QGSJET
interaction model \cite{QGSJET}. 

\begin{table}[h]
\begin{center}
\begin{tabular}{ccccc}
\hline
\hline
 &$a$&$b$&$c$&$d$\\ 
\hline
$e$&0.0041$\pm$0.0004&1.22$\pm$0.005&-0.22$\pm$0.01&6.8$\pm$0.2\\
\hline
$\mu$&0.0079$\pm$0.0003&0.933$\pm$0.001&0.07$\pm$0.002&1.87$\pm$0.05\\
\hline
\hline
\end{tabular}
\caption{Approximation parameters of average EAS electron size
($N_e$) and EAS truncated muon size ($N_{\mu}$) in empirical
formula (4).}
\end{center}
\end{table}

\begin{table}[h]
\begin{center}
\begin{tabular}{ccccc}
\hline
\hline
  &$\alpha$&$\delta$&$\varepsilon$&$\rho$\\ 
\hline
$e$&0.79$\pm$0.01&0.25$\pm$0.003&1.82$\pm$0.05&2.1$\pm$0.1\\
\hline
$\mu$&0.283$\pm$0.003&0.277$\pm$0.003&1.12$\pm$0.05&0.26$\pm$0.1\\
\hline
\hline
\end{tabular}
\caption{Approximation parameters of $\sigma_{e,\mu}$
in empirical formula (4).}
\end{center}
\end{table}

The accuracy of
approximation (4)  for average EAS electron size
$<N_e(E,A,\theta)>$ is less
than 10$\%$ at $3\cdot10^5<E<3\cdot10^8$ GeV, $A\equiv1,\dots56$,
$\theta<32^0$ and observation level 1020 g/cm$^2$. 
The corresponding accuracies of $\sigma_{e,\mu}$ 
and average EAS truncated muon size $N_{\mu}$ are less than
1-2$\%$.\\ 
Changing the variables of kernel functions of Eq. (3) from $N_e$
and $N_{\mu}$ to $x_e$ and $x_\mu$ respectively according to 
\begin{equation}
x_{e,\mu}\equiv A^\delta 
\left(\frac{N_{e,\mu}}{<N_{e,\mu}>}-1\right)+1
\end{equation}
we obtain the following set of Fredholm integral equations

\begin{equation}
\frac{\Delta I(\theta)} {\Delta N_{e,\mu}}=
\int
f_{e,\mu}(E,A)F_{e,\mu}(E,\theta)dE
\end{equation}
where 
\begin{equation}
f_{e,\mu}(E,A)=\sum_{A}
\frac{\partial \Im_{A}}{\partial E}A^{\delta_{e,\mu}-c_{e,\mu}}
\end{equation}
and kernel functions
\begin{equation}
F_{e,\mu}(E,\theta)=
\frac{\partial W(E,\theta)}{\partial x_{e,\mu}}
\frac{1}{a_{e,\mu}E^{b_{e,\mu}}\cos^{d_{e,\mu}}\theta}
\end{equation}
independent of kind of primary nuclei.\\
Examples of distribution functions  $\partial
W/\partial x$ for EAS electron and truncated muon
size spectra are presented Fig.~1,2 respectively.
\begin{figure}[h] 
\begin{center}
\mbox{\epsfig{file=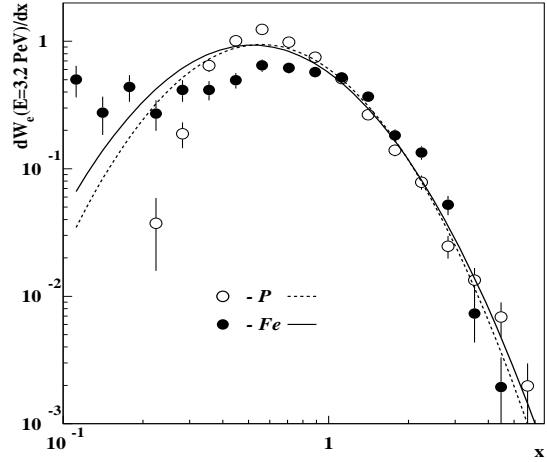,width=8cm,height=7cm}} 
\end{center}
\vspace{-0.5cm} 
\caption{EAS electron size spectra for primary
proton (empty symbols) and iron (filled symbols) nuclei at energy
$E=3.2$ PeV, zenith angle $\theta<18^0$ and observation level 
1020 g/cm$^2$. The lines are the corresponding log-Gaussian
distributions.} 
\end{figure}
These data were obtained by CORSIKA code for primary proton
($A\equiv1$, empty symbols) and iron ($A\equiv56$, filled
symbols) nuclei at 
energy $E=3.2\cdot10^6$ GeV and zenith angle $\theta<18^0$. 
The lines in Fig.~1,2 correspond to log-Gaussian
distributions which were counted based on known values 
of mean $<x>=1$ and variances $\sigma^2_{e,\mu}(E,\theta)$.\\
It is seen, that the distribution functions slightly depend
on a kind of a primary nucleus, especially the right-hand sides of
distributions. It is an important fact, because the
contribution of the left-hand sides of distributions is negligible
small at {\textit{a priori}} steep primary energy spectra
($\partial\Im_{A}/\partial E\sim E^{-3}$).\\
Let the functions $f_e^{(j,k)}(E_i)$ and
$f_{\mu}^{(j,k)}(E_i)$ be the
solutions of the set of Fredholm equations (7) at $i=1,\dots m$
values of primary energies $E_i$, $j=1,\dots n$ kinds of 
primary nuclei and k=1,2,3 zenith angles. Then the values of
unknown
energy spectra ($\partial\Im_{A_j}/\partial E_i$) at a given set 
of ${E_i}$ and different
primary nuclei ($A_j\equiv A_1,\dots A_n$) are determined by the
set
of linear equations
\begin{equation}
<f_{e}^{(j)}(E_i)>_\theta=\sum_{A=A_1}^{A=A_n}
A^{\delta_e-c_e}
\frac{\partial \Im_{A}} {\partial E}{\Big{\arrowvert}}_{E=E_i}
{\ }:{\ } i=1,\dots m
\end{equation}
\begin{equation}
<f_{\mu}^{(j)}(E_i)>_\theta=\sum_{A=A_1}^{A=A_n}
A^{\delta_{\mu}-c_{\mu}}
\frac{\partial \Im_{A}} {\partial E}{\Big{\arrowvert}}_{E=E_i}
{\ }:{\ } i=1,\dots m
\end{equation}
where
$<f_{e,\mu}^{(j)}(E_i)>_\theta=\frac{1}{3}\sum_kf_{e,\mu}^{(j,k)}(E_i)$\\
Evidently, the single solutions of linear equations (9,10)
occur only at $mn\leq2m$, where $mn$ is a number of unknown
$(\partial\Im_{A_j}/\partial E_i)$ and $2m$ is a number of
equations (10,11).\\
\begin{figure}[h]
\begin{center}
\mbox{\epsfig{file=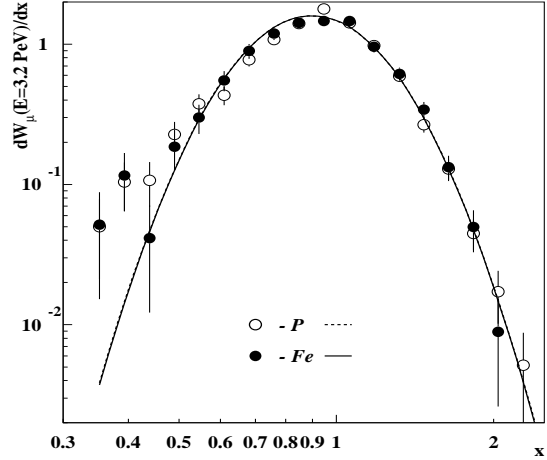,width=8cm,height=7cm}} 
\end{center}
\vspace{-0.5cm} 
\caption{The same as Fig.1 for EAS truncated muon size.}
\end{figure}
Therefore the condition of existence of solutions is $n\leq2$
and the energy spectra \cite{KAS1} obtained based on KASCADE EAS 
electron and truncated muon size
spectra for 4 ($A\equiv1,4,16,56$) primary nuclei no physical
meaning. At the same time
the all-particle spectrum obtained in \cite{KAS1} 
have to be approximately right due to the power index in 
Eq.~(11) $\delta_{\mu}-c_{\mu}\ll1$.\\
It should be noted that unreliability of solutions \cite{KAS1} is
also shown in \cite{Schatz} based on correlation analysis
between the observable electron and truncated muon size spectra 
and unknown energy spectra of primary nuclei.\\ 
\section{Conclusion}
Based on equation (1), the investigation of the EAS inverse problem, 
which is the reconstruction of energy spectra of primary nuclei 
by the observable EAS electron and muon size spectra
at observation level makes sense only using {\emph{a priori}} 
given functions for unknown primary energy spectra with 
given unknown spectral parameters (so called parameterization 
of equation (1)) as it is done in \cite{Glass,STPB,TH,TB}.

\end{document}